\documentclass[10pt,letterpaper]{article}

\usepackage{iccm}
\iccmfinalcopy % Uncomment this line for the final submission 
\usepackage[breaklinks=true]{hyperref}
\usepackage{breakurl}

\usepackage{pslatex}
\usepackage{apacite}
\usepackage{float} % Roger Levy added this and changed figure/table
\usepackage[hybrid,citations,notes,texMathDollars]{markdown}
\usepackage{listings}
\usepackage{amsmath}
\usepackage{mathtools}

\title{Hey Pentti, We Did It!: A Fully Vector-Symbolic Lisp}
 
\author{{\large \bf Eilene Tomkins-Flanagan (eilenetomkinsflanaga@cmail.carleton.ca)}\\
                    {\large \bf Mary Alexandria Kelly (mary.kelly4@carleton.ca)} \\
  Department of Cognitive Science, Carleton University \\
  1125 Colonel By Drive, Ottawa, ON, K1S 5B6, Canada}

\begin{document}

\maketitle

\begin{abstract}
Kanerva (2014) suggested that it would be possible to construct a complete Lisp out of a vector-symbolic architecture. We present the general form of a vector-symbolic representation of the five Lisp elementary functions, lambda expressions, and other auxiliary functions, found in the Lisp 1.5 specification \cite{McCarthy1960}, which is near minimal and sufficient for Turing-completeness. Our specific implementation uses holographic reduced representations \cite{Plate1995}, with a lookup table cleanup memory. Lisp, as all Turing-complete languages, is a Cartesian closed category \cite{Cartesian2024}, unusual in its proximity to the mathematical abstraction. We discuss the mathematics, the purpose, and the significance of demonstrating vector-symbolic architectures' Cartesian-closedness, as well as the importance of explicitly including cleanup memories in the specification of the architecture. 

\end{abstract}

At Clojure/Conj 2023, the conference of the Clojure programming language, \citeA{Meier2023} introduced vector-symbolic architectures to the Clojure community. Her presentation echoed a motif oft heard listening to programmers who discover vector-symbolic architectures (VSAs) for the first time; namely, VSAs' unusual properties and computational niceties are objects of fascination, but it is not immediately obvious what good a vector-symbolic architecture does for the programmer. We present an existence proof that VSAs are completely general computational tools. In technical terms, VSAs can do anything one wants. In pragmatic terms, ``technically anything'' does not answer questions of naturalness and ease of representation. To answer practical questions, VSAs have been used most frequently in representing human cognition, fruitfully in simultaneous localization and mapping (SLAM), and, pertinent to our analysis, promisingly in the syntactic manipulation of neural network states.

In a remark at the end of her talk, Meier mentioned a ``challenge'', issued by Kanerva, in ``one of his papers'', to implement Lisp using exclusively a vector-symbolic architecture representing all the language's expressions. However, the exact words read in the talk as a challenge, ``One could create a `High dimensional computing-Lisp' '', do not seem to have been written by Kanerva. This apparent mistake is not Meier's, as those exact words have been published in \citeA{Neubert2019}, who attribute the enclosed quote to \citeA{Kanerva2014}. While the quote does not appear in Kanerva's paper, the mistake is plausibly a case of miscitation of something said during the associated conference talk, and in any case it is not serious. Kanerva's paper, disappointingly, does not include any challenge, but rather a discussion of how a vector-symbolic Lisp might be implemented, coupled with a loose specification of some of the tools that might be required to do so. We are going to pretend that counts as a challenge, and fully specify a Lisp language in terms of a generic VSA\footnotemark[1].

\footnotetext[1]{Our implementation may be found at \url{https://github.com/eilene-ftf/holis}}

\section{Hold Up, What's a VSA?}

A vector-symbolic architecture is an algebra (i.e., a vector space with a bilinear product),

\begin{enumerate}
\item that is closed under the product $\otimes: V \times V \to V$ (i.e., if $u \otimes v = w$, then $u, v, w \in V$),
\item whose product has an ``approximate inverse'' $\overline{\otimes}$ that, given a product $w$ and one of its operands $u$ or $v$, yields a vector correlated with the other operand,
\item for which there is a dogma for selecting vectors from the space to be treated as atomic ``symbols'', 
\item that is paired with a memory system $\mathcal{M}$ that stores an inventory of known symbols for retrieval after lossy operations (e.g., inversion), that can be recalled from $\mathcal{M}(p)$, and which is appendable $\mathcal{M} \twoheadleftarrow t$, and
\item possesses a measure of the correlation (a.k.a., similarity) of two vectors, $\mathbf{sim}(u, v) \in [-1, 1]$, where $1$ and $-1$ imply that $u, v$ are colinear, $0$ that they are linearly independent.
\end{enumerate}

The $+$ and $\otimes$ operators behave analogously to disjunction and conjunction, or set-theoretic union and intersection. Additionally, VSAs may have an analogue for negation, often the vector rejection on Euclidean space $\mathbf{rej}_v(u)$ \cite{Widdows2015}, and permutations $\Pi$, which are typically used to introduce asymmetry to the product operator, by applying different permutations to the operands. For a detailed review of vector symbolic architectures, see \citeA{Kleyko2022, Kleyko2023}. \citeA{Heddes2023} develop a software library for applying VSAs based on Torch.

In our implementation, we use holographic reduced representations \cite{Plate1995}. They are defined over Euclidean space $\mathbb{R}^n$, and have circular convolution as their product, cosine as their similarity, and atomic symbols sampled from a Gaussian distribution. Our memory system is a lookup table.

Some VSAs (e.g., \citeNP{Kanerva1996}) are not defined over vector spaces per se, or otherwise relax some of the above properties, but behave sufficiently similarly to be used in a similar way. The programmer's choice of VSA comes down to preference and different computational conveniences. For the most part, all VSAs are as good as all others.

Vector-symbolic architectures are an answer to an old tension in cognitive science between the actual machinery of the brain and properties cognition is believed to necessarily possess. Brain states, to the one side, are described in terms of the activity of multiple cell populations, and they exist over a fixed number of cells. Information is typically assumed to be distributed over measured populations, degrading gracefully and uniformly when cells are disabled at random. To the other, \citeA{Fodor1988} made a compelling case that central cognition must have states that behave like discrete symbols, that can be strung together in a combinatorial syntax. But, the traditional tool for representing such syntaxes, computer memory, uses strings of bits for individual symbols, composed by concatenation into ever-longer strings.

With disjunction (sum), conjunction (product), inverse, and similarity operations, plus a cleanup memory, a VSA is sufficient to describe any syntax one could want, to a finite precision. Thus, VSAs appear to satisfy the cognitive scientist's parallel demands for syntax and biomimicry.

\section{Cartejian Clojure and Lisp}

A Cartesian closed category \cite{Cartesian2024} is the mathematical generalization of what it means to compute. It generalizes the equivalence of universal Turing machines (\citeNP{Li2008}, ch. 1) with other definitions of computation, set theory, first-order logic, and, of interest to us, the recursively enumerable languages $\mathbf{RE}$ \cite{Chomsky1955}. All instances of a Cartesian closed category have the preceding equivalences; to say that $C$ is Cartesian closed is also to say that it is Turing-complete. It follows that $C$ can define $\mathbf{RE}$, and, therefore, any syntax, as the language generated by any syntax rules, or grammar, is a subset of that generated by $\mathbf{RE}$.

Categories have two contents: \emph{objects} and \emph{morphisms}\footnotemark[2]. For example, while we normally treat vector spaces as sets of vectors augmented with some functions, they are equally categories that \emph{include} both vectors (their objects) and functions (their morphisms). Cartesian closed categories in particular are useful because they are very simple, and so it is usually easy to demonstrate that a formal system is Cartesian closed.

\footnotetext[2]{Generally, a morphism is any way objects can be related such that, if you have two morphisms $f$, $g$, you can construct $h = f \circ g$ such that $h(a) = f(g(a))$ for some appropriate notion of equivalence.}

In a Cartesian closed category $C$, there is (1) one object, called a terminal object $\mathbf{1}$ (so-named because there is a morphism from every object in the category to it). There is also (2) a product that can compose any two objects, under which $C$ is closed, i.e., if $A$, $B$ are objects in $C$, the product of any objects $A \times B$ in $C$ is also an object in $C$. (3) The functions $A \to B$ on objects in $C$ are together an object in $C$, written $B^A$. (4) A morphism that evaluates functions in $C$, parameterized by objects in $C$, $\mathbf{eval}_C: B^A \times A \to B$ , is in $C$.

These four properties give us four tests for whether some formal system $S$ is Cartesian closed. $S$ must have at least one base data object, and we should be able to transform any expression into it (1). $S$ must have some means to compose arbitrary expressions from its objects, that are data objects still usable by $S$ (2). $S$ must be able to express functions that may map any objects to any others, and those functions must be representable as data objects (3). Finally, we should be able to describe a complete interpreter for $S$, in terms of $S$(4).

Keen readers will have noticed why the above defines computation. We have some base symbols; we may construct sequences of symbols, any length; we can specify any function that transforms sequences to other sequences; and, we can evaluate those functions. That is pretty much a description of a universal Turing machine (see \citeauthor{Li2008}, ch. 1).

\citeA{McCarthy1960} described Lisp for the 1.5th time, giving us the mother document of all subsequent Lisp dialects. Its simplicity will enable us to complete our ``challenge'' without taking up a whole book. To demonstrate that VSAs can compute, we need only implement the five ``elementary'' functions of Lisp 1.5, plus some other functions that can be recursively defined in terms of the others. The elementary functions are: \lstinline{CONS}, \lstinline{CAR}, \lstinline{CDR}, \lstinline{EQ}, and \lstinline{ATOM}. Additionally, there are \lstinline{LAMBDA}, \lstinline{COND}, and \lstinline{LABEL}. \lstinline{LAMBDA} is the most important, as it allows us to define lambda expressions, i.e., arbitrary functions. 

In Lisp 1.5, a tuple is represented in the form \lstinline{(a . b)}, where \lstinline{a} and \lstinline{b} are either atomic symbols (written as alphanumeric sequences) or other tuples. A list \lstinline{(a b c)} is equivalent to the tuple \lstinline{(a . (b . (c . NIL)))}, where \lstinline{NIL} is an atomic symbol that represents the end of a list. Naturally, the singleton \lstinline{(a)} is the tuple \lstinline{(a . NIL)} and the empty list \lstinline{()} is just \lstinline{NIL}. The atomic symbols \lstinline{NIL}, \lstinline{T}, and \lstinline{F} are always defined. We'll define the elementary functions, where lowercase letters are variables that may be any valid Lisp expression:

\begin{lstlisting}
    (CONS a b) = (a . b)
    (CAR (a . b)) = a 
    (CDR (a . b)) = b 
    (EQ a a) = T 
    (EQ a b) = F 
    (ATOM (a . b)) = F 
    (ATOM a) = T
\end{lstlisting}

\noindent The preceding definitions use a pattern-matching format, such that the earlier definition takes precedence. Where the same letter is used for two variables, the variables must be identical. In plain English, \lstinline{CONS} takes any two expressions, and constructs a tuple containing them. \lstinline{CAR} takes a constructed pair, and yields the left element, while \lstinline{CDR} does so with the right. \lstinline{EQ} tests whether two atomic symbols are identical, and is undefined for non-atomic symbols. \lstinline{ATOM} tests whether an expression is atomic. This already seems like very little, but armed with an understanding of Cartesian closure, it can be understood that we don't even need all of Lisp to have a Turing-complete language. We just need \lstinline{CONS} (our product), \lstinline{ATOM} (in case it is not immediately clear, there is a morphism from any expression \lstinline{e} to \lstinline{T} given by \lstinline{(ATOM (ATOM e))}), and \lstinline{LAMBDA} (for functions; the Lisp interpreter evaluates expressions and can be described as a Lisp expression). What makes Lisp remarkable as a point of reference is that there is almost no fat on top of the basic building blocks of a bicartesian closed category; we can describe anything we would like recursively in terms of the basic functions. To wit:

\begin{lstlisting}
((LAMBDA NIL e)) = e
((LAMBDA x NIL) a) = NIL
((LAMBDA x ((CAR x) . e)) a) = (
    LAMBDA (CDR x) 
    (a . ((LAMBDA x e) a)))
((LAMBDA x ((c . d) . e)) a) = (
    LAMBDA (CDR x) (
      ((LAMBDA x (c . d)) a) . 
      ((LAMBDA x e) a)))
((LAMBDA x e) a) = (LAMBDA (CDR x) (
    (CAR e) . ((LAMBDA x (CDR e)) a)))
\end{lstlisting}

\noindent The preceding recursively defines lambda expressions entirely in terms of the Lisp elementary functions, provided that arguments are always curried\footnotemark[3]. The above recursive definition has five cases, where any time \lstinline{LAMBDA} is called, the earliest definition that fits the arguments takes precedence. A lambda expression is a three element list, containing \lstinline{LAMBDA}, a list of parameters \lstinline{x}, and an enclosed expression \lstinline{e}. At base, \lstinline{(LAMBDA x e)} does nothing, but it can be called on one argument \lstinline{a}, which may be \lstinline{NIL}, \lstinline{((LAMBDA x e) a)}, and then it is evaluated over \lstinline{a}, returning either a lambda expression where \lstinline{a} is substituted for all instances of the first parameter, or, if there are no arguments left, the resultant body expression with all substitutions made. Our definition assumes that the value \lstinline{a} is not identical to the names of any parameters, as it can cause incorrect behaviour. An implementation must relabel the parameters of the lambda expression with a novel name, before it is evaluated, in order to ensure that no misbehaviour occurs. In our definition, all lambda expressions are always curried, so a function on three arguments \lstinline{a}, \lstinline{b}, \lstinline{c} is called as \lstinline{(((((LAMBDA x e) a) b) c))}, with the final call being implicitly on the single argument \lstinline{NIL}, as \lstinline{NIL} terminates all lists. The parameters, \lstinline{x}, are a list that is assumed to consist of atomic symbols. \lstinline{LAMBDA} is undefined where elements of \lstinline{x} are nonatomic or duplicate.

\footnotetext[3]{A more Lisp-appropriate definition might have been written such that arguments do not have to be curried, but this version was chosen for ease of presentation.}

\lstinline{COND} implements conditional expressions:

\begin{lstlisting}
    (COND ((T . q) e)) = q
    (COND e) = (COND (CDR e))
\end{lstlisting}

\noindent The way conditionals work is pretty straightforward. We write some implications, and when evaluated, we take the first branch whose condition is satisfied after evaluation.

It is worth noting that this form of recursive definition is useful for its terseness, but it is not proper to LISP 1.5, which would require the use of a \lstinline{DEFINE} pseudo-function to instantiate a function definition. \lstinline{DEFINE} is not one of the elementary functions because it just maps a name to an expression in system memory. Recursive expressions are possible \emph{without} using \lstinline{DEFINE}, so the above effects can be achieved (if not persistently named) by using the \lstinline{LABEL} function. \lstinline{LABEL} is the fundamental tool by which recursion is achieved in the Lisp 1.5 specification, but we choose to omit it due to redundancy.

What we notice in the Lisp 1.5 specification is that there is remarkable inclarity as to what is core to the language and how things are formally defined. Understanding Cartesian closed categories, however, helps to clear up some details. We have chosen to define the parts of Lisp that are elementary, plus lambda expressions, and functions that make programming minimally less painful: \lstinline{QUOTE}, \lstinline{COND}, and \lstinline{DEFINE}. We are now prepared to describe the Lisp VSA.

\section{The Lisp VSA}

The logic of a LISP VSA is straightforward. We are going to map all the elementary functions of Lisp to operations in a vector-symbolic architecture. This proves remarkably straightforward. An interpreter for the Lisp VSA reads a Lisp program and, instead of executing, e.g., \lstinline{CONS}, over two bytes in order to make a two-byte array in memory, it will apply the vector-symbolic \lstinline{CONS} over two vectors in order to create a joint representation of the pair, as a single vector. 

One detail that has not been addressed is how atomic symbols are to be constructed. As the dimension of a space grows, fixing one vector and choosing another vector arbitrarily, the expected value of their similarity goes to 0, and vectors with nonnegligible similarity are exceedingly rare \cite{Kanien1993}. In holographic reduced representations \cite{Plate1995}, we sample vectors on $\mathbb{R}^n$ from a normal distribution, with $\mu = 0$ and $\sigma = \frac{1}{\sqrt{n}}$, producing vectors $v$ with $\mathbb{E}[||v||] = 1$. Thus, we can have many more base symbols than dimensions in the space, all \emph{nearly} linearly independent\footnotemark[4], which would not be the case if they were truly orthogonal. For Euclidean vectors, it typical to see $n \in \{2^k, k \in [6, 12]\}$.

\footnotetext[4]{We can calculate the expected variance of pairwise $\mathbf{sim}(a, b)$ for an arbitrary overcomplete basis $B$ (i.e., a finite sample of $\mathbb{R}^n$ where $|B| > n$) with $a \neq b \in B \subset \mathbb{R}^n$ exactly, but that calculation is outside the scope of this paper. A commonly used ``margin of safety'' expects $\mathbf{sim}(a, b) \in (-0.2, 0.2)$, but for $n \ge 512$ the actual expected variance is much smaller, even for large $|B|$.}

\citeA{Kanerva2014} suggested representing lists by permuting one operand then adding the two operands together. By keeping a fixed permutation in memory, the united representation is most similar to its unpermuted operand by default, and then, by applying the inverse permutation, winds up most similar to its permuted operand, with fixed permutation $\Pi$. 

$$
\mathbf{cons}(a, b) = a + \Pi(b)
$$

$$
\mathbf{car}(c) = \mathcal{M}(c)
$$

$$
\mathbf{cdr}(c) = \mathcal{M}(\Pi^{-1}(c))
$$

\noindent This method has some flaws if we care about retrieval, however. Taking advantage of the property just described, either operand can be retrieved from a tuple very simply. But, problematically, if one wishes to make a list of arbitrarily many elements, one needs to store sublists in memory. Once one stores a list in memory, however, the vector to which that list is most similar is itself. It becomes necessary to do something to tag \emph{both} operands, such that each operand and their disjunction may be reliably distinguished in memory.

Permutation is also a redundant operation, if we are implementing a Lisp. Although it is often used to make the conjunction operator $\otimes$ asymmetric, this behaviour is not necessary if we are implementing a Cartesian closed category, as the role of taking an operation that builds a joint representation of two operands, and making it asymmetrical, is already satisfied in the VSA algebra. In the Lisp VSA, the product's job is making  a combined representation of two objects that is dissimilar to either, but both remain retrievable, using their pair and a cleanup memory, which is exactly what permutation is doing in Kanerva. As such, we will leave out the permutation operator and work just with $+, \otimes, \overline{\otimes}, \mathbf{sim}, \mathcal{M}$, 

We define additional operators for convenience. $\mathcal{F}$ is another cleanup memory, that separates global function bindings from the inventory of retrievable expressions. $\nu(v) = \frac{v}{||v||}$ divides $v$ by its magnitude, such that $\mathbf{sim}(\nu(v), v) = 1$, $||\nu(v)|| = 1$. $\oplus$ is a variant addition operator that saturates on both an upper and lower threshold, respectively $\theta_{\uparrow}$, $\theta_{\downarrow}$:
%\end{markdown}

\begin{align}
\begin{split}
a \oplus b ={}& (\mathbb{E}[||a||] > \theta_{\uparrow}) a \\
            & + (\mathbb{E}[||a||] < \theta_{\downarrow}) b \\
            & + (\theta_{\downarrow} \le \mathbb{E}[||a||] \le \theta_{\uparrow})\;\nu(a + b)
\end{split}
\end{align}

%\begin{markdown}
\noindent Because $\oplus$ is defined using three mutually exclusive cases, the operands $a$ and $b$ can be lazily evaluated, such that the values are only computed if they are needed. Defined assuming lazy evaluation, $\oplus$ is a useful operator for writing recursive definitions. Setting $\theta_{\uparrow}$ and $\theta_{\downarrow}$ respectively a little under $1$ and a little over $0$, and the expected magnitude of all vectors is $1$, expressions written with $\oplus$ are meant to be read as evaluating the left operand if a test multiplying it succeeds (the test yields a scalar value $\alpha = \mathbb{E}[||a||] > \theta_{\uparrow}$), and evaluating the right operand if it fails ($\alpha < \theta_{\downarrow}$). Several additional atomic expressions are used in the preceding definitions, notably $L$ and $R$, which mark the left hand and right hand sides of a tuple, as well as $\varphi$, which marks that a vector is nonatomic, and $\rho$, which marks that a lambda expression has been relabeled such that its parameters may not have the same names as possible values of their arguments.

$\mathbf{f}(a)$ marks a call to programmer-defined function $\mathbf{f}$, and requires some special treatment, as $\mathbf{cons}(\mathbf{f}, a)$ is equivalent in our notation to $\mathbf{f}(a)$. What the notation means is that the interpreter should leave the list including $\mathbf{f}$ as an unevaluated expression if $\mathbf{f}$ is not in the function namespace.

Below are the definitions of the Lisp VSA. Single unbolded lowercase letters refer to variables that may contain arbitrary Lisp expressions, but typically they are expected to be of a certain form and lead to undefined behaviour when not of that form. Bolded words and letters refer to function names, and are expected to always be atomic. Functions in general are called by simply using their name, and so all function calls are marked by atomic symbols at the head of a list of arguments, except in the case of lambda expressions, which are three-element lists. Lisp expressions of the form \lstinline{(F a b ...)} are translated to our notation as $\mathbf{f}(a, b, ..., t)$, where the last element $t$ is always the tail of the list of arguments. Recalling that lists are recursively nested tuples, \lstinline{(F a b ...)} is equivalent to \lstinline{(F . (a . (b ...)))}, and likewise $\mathbf{f}(a)(b)... = \mathbf{f}(\mathbf{cons}(a, \mathbf{cons}(b, ...))) = \mathbf{cons}(\mathbf{f}, (\mathbf{cons}(a, \mathbf{cons}(b, ...)))$ in our notation. Programmer-defined functions are always fully curried. The special case of \lstinline{(F)} is, following the definition of lists, equivalent to $\mathbf{f}(\mathrm{NIL})$. Therefore, we never technically have functions on no arguments. Here are the definitions:
%\end{markdown}

\begin{align}
\begin{split}
\mathbf{cons}(a, b, \_) :={}& \nu(L \otimes a + R \otimes b + \varphi) \; | \; \mathcal{M} \leftarrow\mathllap{+} a, b
\end{split}
\end{align}
\begin{align}
\begin{split}
\mathbf{car}(a) :={}& \mathcal{M}(L \overline{\otimes} a) 
\end{split}
\end{align}
\begin{align}
\begin{split}
\mathbf{cdr}(a) :={}& \mathcal{M}(R \overline{\otimes} a)
\end{split}
\end{align}
\begin{align}
\begin{split}
\mathbf{eq}(a, b, \_) :={}& \mathbf{sim}(a, b) T + (1 - \mathbf{sim}(a, b)) F
\end{split}
\end{align}
\begin{align}
\begin{split}
\mathbf{atom}(a, n) :={}& \mathbf{sim}(n, \mathrm{NIL})\mathcal{M}(\mathbf{sim}(a, \varphi) F \\
                        & \qquad+ (2\theta_{\downarrow} - \mathbf{sim}(a, \varphi))^+ T) \\
                        & + (2\theta_{\downarrow} - \mathbf{sim}(n, \mathrm{NIL}))^+F
\end{split}
\end{align}
\begin{align}
\begin{split}
\mathbf{define}(a, e, \_) :={}& * \; | \; \mathcal{F} \leftarrow\mathllap{+} \,\mathbf{cons}(a, e)
\end{split}
\end{align}
\begin{align}
\begin{split}
\mathbf{cond}(r) :={}& \mathbf{sim}(\mathbf{car}(\mathbf{car}(r)), T) \mathbf{cdr}(\mathbf{car}(r))\\ 
                     & \oplus \mathbf{cond}(\mathbf{cdr}(r))
\end{split}
\end{align}
\begin{align}
\begin{split}
(\mathbf{\lambda}(x, e))(a) + v :={}& (\mathbf{sim}(\mathbf{\lambda}(x, e) + v, \rho) < \theta_{\downarrow}) \\
                    & \qquad(\mathbf{\lambda}(\mathbf{relabel}((x, e))) + \rho)(a) \\
                    & \oplus \mathbf{sim}(x, \mathrm{NIL}) e \\
                    & \oplus \mathbf{sim}(e, \mathrm{NIL}) \mathrm{NIL} \\
                    & \oplus \mathbf{\lambda}(\mathbf{cdr}(x), \mathbf{\lambda s}(x, e)(a))
\end{split}
\end{align}
\begin{align}
\begin{split}
(\mathbf{\lambda s}(x, e))(a):={}& \mathbf{sim}(x, \mathrm{NIL}) e \\
                    & \oplus \mathbf{sim} (e, \mathrm{NIL}) \mathrm{NIL} \\
                    & \oplus \mathbf{sim}(\mathbf{car}(x), e) \mathbf{car}(a) \\
                    & \oplus \mathbf{sim}(\mathbf{atom}(e), T) e \\
                    & \oplus \mathbf{sim}(\mathbf{car}(x), \mathbf{car}(e)) \\
                    & \qquad \mathbf{cons}(\mathbf{car}(a), \mathbf{\lambda s}(x, \mathbf{cdr}(e))(a)) \\
                    & \oplus \mathbf{sim}(\mathbf{atom}(\mathbf{car}(e)), F) \\
                    & \qquad \mathbf{cons}( \\
                    & \qquad\quad (\mathbf{\lambda s}(x, \mathbf{car}(e)))(a), \\
                    & \qquad\quad(\mathbf{\lambda s}(x, \mathbf{cdr}(e)))(a) \\
                    & \qquad) \\
                    & \oplus \mathbf{cons}(\mathbf{car}(e), (\mathbf{\lambda s}(x, \mathbf{cdr}(e)))(a))
\end{split}
\end{align}
\begin{align}
\begin{split}
\mathbf{f}(a) :={}& \mathbf{sim}(\mathbf{f}, \mathbf{car}  (\mathcal{F}(L \otimes \mathbf{f}))) \\ 
                    & \quad \mathbf{cons}(\mathbf{cdr}(\mathcal{F}(L \otimes \mathbf{f})), a) \\
                    &\oplus \mathbf{cons}(\mathbf{f}, a)
\end{split}
\end{align}

%\begin{markdown}

With some minor modifications due to simplifications of the specification, the above definitions can be used to implement the Lisp 1.5 interpreter (\citeauthor{McCarthy1960}, Appendix B). The function $\mathbf{relabel}$ generates a new list of parameters $y$ consisting of random atoms, and substitutes all occurrences of each parameter $x_1, x_2, ...$ in $e$ for the corresponding $y_1, y_2, ...$, then returns $(y, e)$. Its definition is omitted in the above because it would significantly complicate the presentation.

What is particularly notable in the above definitions is the frequency and fundamentalness of the use of cleanup memories. Every VSA has a cleanup memory, but usually, the cleanup memory relies on a big matrix $M$ that stores every known symbol as an approximately unit length row vector. Thus, on $\mathbb{R}^n$, the cleanup memory is $\mathcal{M}(p) = \mathbf{hardmax}(M^Tp)M$, where $p$ is a probe vector to be ``cleaned up'', by retrieving its nearest neighbour from memory. $\mathbf{hardmax}$ is defined to be equal to $1$ on the dimension its input is maximal, and $0$ everywhere else. Because of the historic difficulty of implementing both time- and space-efficient cleanup memories, and a low appraisal of the biological significance of ``memory is a big lookup table and you test every entry in order to retrieve the one you want'', the choice of cleanup memory being used by any given VSA is an embarrassment one typically glosses over (e.g. while \citeNP{Kanerva2014} discusses cleanup memories, they are not explicit in his algebraic notation, and are left as a black box in his diagrams). We emphasize the explicit notation of cleanup memories, because they are essential for achieving features of VSAs in frequent day-to-day use, because different cleanup memories have distinctive computational properties that may fit some applications better than others, and because memories with certain computational properties are essential to achieving Turing-completeness in our Lisp.

\citeA{Kanerva2014}, following \citeA{Eliasmith2012}, refers to the vectors of a VSA as ``pointers''. That is because, partly, a probe in the memory can be taken to ``reference'' its nearest neighbour; a trace can be looked up using any of the points in space near it. Traditional memory pointers similarly ``probe'' memory, though, in general, what is at the probed memory location need not have high mutual information with the probe. Variant cleanup memories can also be defined that are similarly \emph{heteroassociative}, making probes much more like traditional pointers. One glaring flaw appears in the pointer analogy, however: Where $d$ is the number of stored traces, retrieval from computer memory is $O(\mathrm{log} (d))$, if $d$ is close to the total available space ($O(1)$ if significantly less). Probing a traditional cleanup memory is at least $O(d)$. In fact, because probing memory \emph{requires} traversing all stored traces to test for similarity, the lookup is also at least $\Omega(d)$. It is not an issue of principle versus practice either; because VSAs use vectors of extremely high dimensionality, comparisons take a long time, and because one is often storing thousands or millions of vectors in memory for practical applications, one is really getting one's $n$'s worth of comparisons in.

\section{So What is to be Done?}

\citeauthor{Neubert2019} make a second apparent misattribution to \citeA{Kanerva2014}, which is also fruitful to pretend was written as attributed. They suggest the possibility of another type of cleanup memory: an attractor neural network. In such networks, information is often (though not always) distributed over the network's weights, which makes them robust to noise or damage, as the vector representations of VSAs are robust. Attractor networks feature interacting cells converging to stable patterns over time, a tantalizingly brain-like property. However, most attractor networks in use are no more time or memory efficient than a big matrix. The Hopfield network (\citeNP{Hopfield1982}, made continuous in \citeNP{Hopfield1984}) has a storage capacity of $O(n)$ with respect to its input dimensions. Hopfield networks work almost \emph{exactly like} the big matrix format (with some care paid to representation choice, they may produce approximately equal results, \citeNP{Kelly2017}), with a different \emph{activation function} (above, our activation function was $\mathbf{hardmax}$) and the proviso that the network's outputs may be fed back into it, until it converges\footnotemark[5].

\footnotetext[5]{Hopfield called for updating neuron activations at random, but both bulk and random updates converge to the same outcomes.}

Another appealing option is to use the match networks of \citeA{Grossberg2021}, as they've seen some success in modelling human brains, and also claim to solve retroactive interference. Unfortunately, they also look like the big matrix approach of before\footnotemark[6], and they eliminate retroactive interference by ``gating'' gradient descent, with a function that updates only on one row at a time, prohibiting the storage of more than $O(n)$ traces or retrieving them in less than $\Omega(d)$ time, if traces are assumed to have low mutual information.

\footnotetext[6]{Converging to $\mathcal{M}(p) = \tau(\sigma(M^Tp)M + p)$, where $\sigma$ behaves similarly to $\mathbf{hardmax}$, and $\tau(v) = \varsigma(\frac{v}{||v||_1})$ with logistic function $\varsigma$.}

If we relax the requirement that traces be near-orthogonal, better results may be obtained. \citeA{Orobia2022} use a continuous variant of MINERVA2 \cite{Hintzman1984} as the memory system of a reinforcement learning agent. MINERVA2 also resembles the ``big matrix'' memory: $\mathcal{M}(p) = (M^Tp)^\rho M$ where $\rho$ is an odd integer power. $\rho$ can be allowed to be a real number using the variant equation $\mathcal{M}(p) = \mathbf{sgn}(\xi)(\mathbf{sgn}(\xi) \xi)^\rho M$ where $\xi = M^Tp$. Traces are still inserted row-wise, but \citeauthor{Orobia2022} do not expect to retrieve traces exactly as-stored, and rather interpolate between stored traces using probes similar to several of them. They also employ a forgetting mechanism: when information is unused, it fades out of memory. Thus, the size of memory is capped, without running out of space. Their system is not strictly vector-symbolic; there is no syntactic manipulation. But, if atomic symbols are allowed to be correlated and we permit forgetting, similarly advantageous properties may be usable. One reason to specify the exact cleanup memory used in one's VSA is that its space, timing, and information loss characteristics are very relevant topics for study. Different tradeoffs of characteristics might significantly affect the behaviour of the VSA in a specific use-case.

For our vector-symbolic Lisp, MINERVA2 is inadequate, at least without significantly modifying the specification. Let us reflect on the general form of the cleanup memories we have looked at: $\mathcal{M}(p) = \tau(\sigma(\beta M^Tp) M + q)$ with activation function $\sigma$, normalization function $\tau$, scalar constant $\beta$, and some added factor $q$. In most of the preceding cases, $\tau$ has been the identity, $\beta = 1$ and $q = 0$. $M$ is an $m \times n$ matrix, where $m$ is typically $O(d)$, each row storing one $n$-dimensional trace. Ideally, we want $M$ to distribute information about retrievable traces, as in the case of Hopfield networks and MINERVA2; we want to retrieve traces exactly as-stored, as in the big matrix case; we want to store a number of traces that is superlinear relative to the input dimension $n$, both for the sake of having a cleanup memory with a great capacity, and for improving the memory's timing characteristics. Capacity is important, because our Lisp relies so heavily on memory to allow for a program to be written with arbitrary functions, and because the set of functions on a $S$ is the powerset $\mathcal{P}(S)$; if arbitrary programs are allowed, we need to be able to store and retrieve arbitrary sequence-to-sequence maps, requiring a memory system that is at least exponential in storage capacity with respect to the input dimension. Just one candidate cleanup memory fits the bill: the modern Hopfield Network (MHN; \citeNP{Ramsauer2021}).

Mathematically demonstrating an exponential storage capacity and therefore $O(\mathrm{log} (d))$ lookup time, \citeauthor{Ramsauer2021} describe a neural network with a familiar form:

$$
\mathcal{M}(p) = \mathbf{softmax}(\beta \; M^T p) M
$$

\noindent Information is distributed because stored traces are encoded in $M$ using gradient descent (no gating). Special cases are the big matrix variant (equal to the limit of the MHN equation as $\beta \to \infty$) and a linear memory system (with $\beta = 0$). If $\beta$ is allowed to be a function of $\xi$, then MINERVA2 is also a special case (after normalization) with $\beta = \frac{\rho \log(x)}{x}$ (demonstrable algebraically), but this is not a very useful example. The special cases serve to indicate that the capacity of the network is sensitive to the choice of $\beta$. Given an arbitrary program, then, some amount of tuning is necessary to make the MHN store what needs to be stored.

However, MHNs obtain an exponential capacity by way of encoding by gradient descent, so, while capacity is large and retrievals are $O(1)$ as a function of the number of stored traces, encoding is $O(n)$, and requires a backup of all stored traces, if retroactive interference is to be avoided. As it is possible for many traces to be stored at runtime in the VSA Lisp, MHNs present significant drawbacks for us, so long as they depend on a gradient descent learning rule.

An improvement might be made to the MHN by using a variant of Grossberg's gating function. The activation function of the match network is \emph{essentially} similar to that of the MHN, albeit using another function $a$ in place of $\mathbf{softmax}$. The most significant distinction is in the update rule, where Grossberg uses a modified version of single-layer gradient descent, where $\nabla_M L$ is the partial gradient of some \emph{loss} or \emph{error} function $L$ with respect to $M$, and $\eta$ is a learning rate:

$$
R_G = M_{t + 1} = M_t - \eta \; \mathbf{hardmax}(M^T p) \nabla_W L
$$

\noindent The $\mathbf{hardmax}$ function is of the same dimension as its input, and is 1 where the input dimension is greatest and 0 on all others. Intuitively, only one row of the memory matrix $M$ can be updated at a time in this fashion. Observing that the definition of gradient descent is:

$$
R_C = M_{t + 1} = M_t - \eta \nabla_W L
$$

\noindent We may observe that Grossberg's rule and the original gradient descent rule are two limit cases of \emph{one} rule, with multiplicative parameter $\alpha$:

$$
R_E = M_{t + 1} = M_t - \eta \, \alpha \; \mathbf{softmax}(\gamma \; M^T p) \nabla_W L
$$

\noindent Respectively, $R_C = \lim_{\gamma \to 0} R_E[\alpha:=n]$, and $R_G = \lim_{\gamma \to \inf} R_E[\alpha := 1]$. As $R_G$ minimizes interference between stored patterns, but also minimizes capacity, while $R_C$ maximizes both, the continuum of $R_E$ suggests a noninterference-capacity tradeoff varying with the parameter $\gamma$. Although the question of whether a large capacity can be preserved while interference remains sufficiently low for our purposes remains open, we are optimistic.

\section{And Whatfor This Lisp?}

The behaviour of vector-symbolic architectures is very sensitive to the choice of cleanup memory. While memory characteristics are not typically used to demonstrate Turing-completeness, Turing-completeness makes sufficiently great demands of memory that reasonable performance requires specific memory characteristics. As the computing applications of VSAs expand, studying these characteristics and making good tradeoffs will be very important.

One application that stands out is suggested by \citeA{Tamkin2023}, who trained a transformer network to exhibit states that were decomposable into ``monosemantic'' vectors $S$. The semantic content of the network's output was manipulated by adding or subtracting features drawn from $S$. As such, transformer states may be made to behave as additive compositions of atomic vector symbols, of the sort syntactically composable by VSAs.

Creating a vector-symbolic Lisp has been alluded to a few times, in particular by \citeA{Kanerva2014}, \citeA{Smolensky1990}, and \citeA{Legendre1990}. The appeal is obvious to cognitive scientists and explicit in \citeauthor{Smolensky1990}: We think that brain states have syntax, and we know information is distributed over them. VSAs are a means to express syntax in terms that may describe brain states, and Lisp instantiates the most general class of syntaxes. Respecting actual neural networks, \citeA{Chen2020} and \citeA{Smolensky2022} have put syntactic manipulation of network states into practice, with promising results. 

Traditional cognitive architectures, such as ACT-R, describe memory states syntactically \cite{Stewart2006}, and take actions according to rules that are sensitive to syntactic features. Such states have already been described in terms of a VSA \cite{Kelly2020}, and, taking into account their rules and memory systems, these cognitive architectures are already Turing-complete. However, it is uncommon for these cognitive architectures to treat memory states as arbitrary programs, and attempt to evaluate them. It is notable that these architectures directly descend from an attempt to describe artificial general intelligence \cite{Newell1980}. A key demand of general intelligence in Newell is for the general intelligent agent to ``behave as an (almost) arbitrary function of the environment'' (p. 139). That statement requires the \emph{procedural memory} (i.e., the memory system that encodes behaviours) to be Turing-complete, but it is not immediately obvious that the mental representations procedural memory \emph{operates over} must encode programs as well. One property of Turing-completeness is that, in a Turing-complete system, it must be possible to define a program (in our case, a behaviour) that interprets memory objects as programs and attempts to execute them. The ability to do so follows from the capacity of any Turing-complete machine to simulate any other.

\citeA{Turing1950}'s universal machine was originally a description of the sorts of things people do in their head, manipulating either their memory or a piece of paper. As a model of \emph{human} computers, the universal machine could, as humans do, store methods for computing some functions in its memory, and interpret those methods as programs to compute results. The notion of interpretation is particularly salient for us. Above, we described (as one does in Lisp) lambda expressions as lists, and defined a behaviour that interprets them as functions and causes them to be evaluated. An arbitrary program may be described in this way.

Arbitrary computation is not part of the daily life of most people. From a modelling perspective, interpreting the states of memory as programs does not seem, at first blush, like it would be very useful, as most tasks we may want to model do not seem to require Turing-complete representations. \citeA{Dehaene2022} suggest, however, that mental programs are in play in a way we may not even be cognizant of. They observe that reaction times, error rates, and some neuroimaging data in humans, but not in other species, in a visual pattern-recall task, are better predicted by the \emph{algorithmic information} of those patterns, than the \emph{classical information}. In algorithmic information theory, the information content of a string of symbols drawn from a finite set $x$ is the length $\ell$ of the shortest possible program $p$ that prints $x$ and then halts. \citeauthor{Dehaene2022} theorize that mental representations must, therefore, be the shortest length programs that \emph{reconstruct} whatever they are representing (here, a visual pattern).

\citeauthor{Dehaene2022} suggest, without much in the way of explanation, that representations being mental programs has something to do with general intelligence (their paper is subtitled ``a hypothesis about human singularity'', referencing the ``technological singularity'' that is speculated to follow the discovery of artificial general intelligence). The meaning of their intertextual indulgence is not immediately apparent, but it may be an oblique reference to the only formal description of artificial general intelligence that exists, AIXI \cite{Hutter2000}. The name, AIXI, stands for Artificial Intelligence crossed with Induction. Hutter uses a Solomonoff induction, a form of Bayesian induction that makes use of algorithmic information \cite[ch. 5]{Li2008}, to describe a universally optimal reinforcement learning agent. While Hutter's model is not computable, and intractable even in restricted cases, it is a useful formalization because it proves that the agent will learn to behave rationally, and will do so faster than any other without significant domain knowledge, supplying a well-defined upper bound on rational behaviour (\citeNP{Newell1980} wrote that AGI should ``exhibit rational, i.e. effective adaptive behaviour'', p. 139, and intends rationality to be ``goal-directed'', p. 171) in pursuit of a goal encoded in some reward function \cite[p. 53, statement of the \emph{reward hypothesis}, but we note that it's not a hypothesis]{Sutton2018}.

A fundamental intuition in AIXI is that observations of the world can be represented as finite strings $x$. Based on what the agent has seen of the world, it assigns a measure of the likelihood of observing any sequence of events $x$ with the following function $\xi_U$ (the variables in this section are reproduced from \citeNP{Hutter2005}, ch. 1, and are unrelated to prior sections):

$$
\xi_U(x) := \sum_{\nu \in \mathcal{M}_U} 2^{-K(\nu)}\nu(x)
$$

\noindent $\mathcal{M}_U$ in the preceding equation is the set of functions $\nu$ that compute the probability of all observed sequences. The function $\mu \in \mathcal{M}_U$ is the ``true'', but unknown, probability distribution of sequences of events. $K$ is the function representing the algorithmic information of the distribution function $\nu$. Intuitively, $K$ represents the length of the shortest function that produces all of the predicted probabilities $\nu$ does. It is defined:

$$
K(\nu) := \min_p \left\{ \ell(p) \; : \; U(p) = \nu \right\}
$$

\noindent Above, $\ell$ computes the length of a program $p$ that runs on a universal Turing machine $U$ to compute $\nu$. $K$ is the length of the shortest program that computes $\nu$. By using an exponential decay of $K$, the above function, $\xi_U$ is weighted by the brevity, or parsimony, of programs that may produce the observations that have been made. 

There are three infinities in the above equations: $\mathcal{M}_U$ is infinitely large, the programs $p \in P$ are infinitely numerous, and finding the ones that compute $x$ runs into the halting problem, while the set $X$ containing all possible sequences of events $x$ is infinitely large. All of these must be further constrained to produce a computable, let alone efficient, model.

As we are already trying to find the shortest $p*$ that generate the same predictions as $\nu \in \mathcal{M}_U$, we can pretend that we are just considering $P$. Instead of finding out what all possible programs compute (this can only be done by executing them) and averaging over the lot, we will try to find close-to-optimal programs to represent our $\nu$, and we will try to only consider the $\nu$ that are likely to be $\mu$. These are two search problems, defined over the set $P$. On the one hand, given an equivalence class defined by what programs $p_1, p_2, p_3... \cong \nu$, perform the same computation as $\nu$, we want to find or converge on $p* = \mathrm{argmin}_p \left\{ \ell(p) \; : \; U(p) = \nu \right\}$ for each $\nu$ in consideration, and on the other hand, given our observations of the world so far, we want to remove from consideration $\nu$ that are unlikely to be predictive, or assign a low probability to actually observed $x$. 

General intelligence implies some capacity to represent and solve problems (or, figure out how to satisfy goals) in arbitrary domains (or environments), a feature that was unfortunately omitted from the text of Newell's original discussion. That capacity is found in the AIXI model, and AIXI specifies an upper bound on how efficient this problem-solving can be. General intelligence is derived from observation of what humans do, but much more loosely than Turing's derivation of universal computation, so it may not be possible to say so convincingly that humans are general intelligent. But, \emph{were} that the case, and \emph{were} humans also close to \emph{optimally} rational, up to the bounds permitted by homeostasis, then we would expect human mental representations to follow such a procedure, approximating the kinds of representations we would find in AIXI. That being the case, humans would \emph{have} to store mental programs, search for the best ones to represent their entertained theories $\nu$ about the world they are observing, and select the best $\nu$ that explain their observations.

Beyond the strict necessity that brain states should therefore be interpretable as programs, there is also the significant matter of the \emph{benefits} of encoding programs on a vector space. One of those benefits may seem to be machine learning using neural networks, though particular consideration must be paid to designing encodings such that neural networks can take advantage of them. In the search for efficient encodings of a $\nu$, for instance, we defined an equivalence class of programs $p_1, p_2, p_3... \cong \nu$, given by the identity $\forall x \in X. U(p_i)(x) = \nu(x)$. For a neural network to efficiently traverse this class in search of the simplest $p*$, programs with similar size and structure should be correlated. Likewise, when selecting theories $\nu$ likely to be $\mu$, we want to be able to smoothly traverse theories by way of their similar predictions, so programs in similar equivalence classes should also be correlated.

If humans are taken to be general intelligent, then it is strictly necessary for mental representations to be interpretable as programs. To that end, it is only necessary to describe rules that treat certain states of memory as lambda expressions and evaluate them (as we have done). Then, memory may encode arbitrary programs, although, based on the behaviour of our own interpreter\footnotemark{1}, our simple design is not the most efficient that can be achieved. There are candidate memory systems that likely outperform what we have tried so far, there are encodings that take better advantage of the features of continuous space to make programs efficiently \emph{learnable}, and there are likely encodings that represent functions in a manner more amenable to efficient evaluation.

Taking programs to be the sorts of representation that can be used in the most general kind of problem-solving, and taking \citeA{Newell1980}'s commitment that representations provide ``distal access'' to and control of the represented object, one thing that falls in this maximally general class is the model producing and using representations. It is striking that, in several leading functional theories of consciousness, a necessary feature is one's ability to pursue one's goals by reading and manipulating of one's own internal state \cite[p.5, Table 1; RPT-1, 2, HOT-2, 3, AST-1, AE-1]{Butlin2023}. By identifying atomic states, and training a network to represent compositions of states, an arbitrary syntax can be defined over a network's state space, though not all networks can be trained to make all syntaxes useful. As we have seen, it takes specific features to make a syntax learnable, and only some networks can efficiently traverse through an encoded structure.

But, assuming useful syntaxes are possible, then the states of some networks can be encoded in some sequence of steps that traverse the space from an initial point. Such sequences may be generable by some function of the kind that the network being traversed is capable of learning. Sequence-to-sequence models, if they generate structured internal representations, may therefore be capable of reading and writing to the states that control their behaviour, in a rational, goal-directed fashion. Because VSAs are provably Turing-complete, there are no limits to how learnable programs can subject network states to syntactic manipulation. If any of the noted theories surveyed in \citeA{Butlin2023} are correct in requiring such auto-manipulation, then \emph{vector-symbolic architectures might even be the gateway to machine consciousness}.

\bibliographystyle{apacite}

\setlength{\bibleftmargin}{.125in}
\setlength{\bibindent}{-\bibleftmargin}

\bibliography{ref}

\end{document}